\documentclass[reprint,nofootinbib,amsmath,amssymb,aps,floatfix,superscriptaddress,twocolumn]{revtex4-2}
\usepackage{graphicx}
\usepackage{dcolumn}
\usepackage[colorlinks,linkcolor=magenta,anchorcolor=cyan,citecolor=blue]{hyperref}
\usepackage{bm}
\usepackage[multiple]{footmisc}

\begin{document}
\title{Observational Constrains on the Sgr A$^*$ Black Hole Immersed in a Dark Matter Halo: Shadow and S2 Star Orbit}


\author{Zhen Li} 
\email{zhen.li@just.edu.cn}
\affiliation{School of Science, Jiangsu University of Science and Technology, Zhenjiang 212100, China}

\begin{abstract}
It is widely believed that Sgr A$^*$, located at the center of our Galaxy, is a supermassive black hole. Recent observations of its shadow and long-term monitoring of the S2 star have provided compelling evidence supporting this hypothesis. These observational advancements also offer valuable opportunities to explore the physical properties of the black hole and its surrounding environment. Since a dark matter halo is expected to exist in the Milky Way and around Sgr A$^*$, investigating the behavior of the Galactic Center black hole embedded in such a halo provides a crucial means to simultaneously probe both black hole physics and dark matter properties. In this work, We develop a black hole metric that incorporates a generalized double power law dark matter halo, and analyze the corresponding null and timelike geodesics to investigate how the halo parameters affect the black hole shadow and the motion of the S2 star. Furthermore, by comparing our theoretical predictions with observational data of the shadow and the S2 orbit, we constrained the dark matter halo parameters. The results of this study provide both theoretical and phenomenological insights into the nature of Sgr A$^*$ and the distribution of dark matter in our Galaxy.
\end{abstract}
\maketitle

\section{Introduction}

Black holes are among the most fascinating objects predicted by Einstein’s theory of gravity and remain at the forefront of modern physics. Groundbreaking observations, such as the detection of gravitational waves by LIGO \cite{gw1,gw2,gw3} and the imaging of black hole shadows by the Event Horizon Telescope (EHT) \cite{shadow1}, have opened new windows into this field. Recently, the shadow of the supermassive black hole at the center of our Galaxy, known as Sgr A$^*$, has also been observed \cite{shadow2}. The black hole shadow arises from the trajectories of photons in the strong gravitational field near the event horizon \cite{len1, len2, len3}, providing a novel probe of the physics in this extreme regime. 

Meanwhile, continuous observations of stellar motions around the Galactic Center have marked another milestone in astrophysics. In particular, the S2 star serves as an ideal object for studying the dynamics near the Galactic Center, owing to its long-term, high-precision measurements of position and radial velocity \cite{ss1,ss2, pdata1,pdata2,pdata3,pdata4,vdata1,vdata2}. Observational data of the S2 star have been used to study various aspects, such as estimating the distance and mass of Sgr~A$^*$ \cite{s2app1}, constraining dark matter spikes \cite{dmbh17}, investigating the orbital precession of S2 in modified gravity \cite{s2app3}, probing the mass distribution in the Galactic Center \cite{mpro}, examining scalar clouds around Sgr~A$^*$ \cite{s2app5}, and testing quantum-corrected black holes \cite{s2app6}. 

The black hole shadow and the motion of the S2 star are closely related to null-like and time-like geodesics, respectively. Both are strongly influenced by the spacetime geometry and the surrounding matter distribution, leading to observable effects. Therefore, they serve as powerful tools for probing the nature of black holes and their environments.

Dark matter is another mysterious and fascinating topic in modern physics. It neither emits nor reflects light, making it extremely difficult to detect directly in experiments. Dark matter is generally assumed to be collisionless, interacting with ordinary matter only through gravity. Its presence is inferred indirectly from phenomena such as gravitational lensing and galaxy rotation curves \cite{dm1, dm2, dm3}. At present, N-body simulations provide the most effective means of studying the large-scale distribution of dark matter in the Universe \cite{dms1,dms2}. These simulations suggest that dark matter forms halos surrounding galaxies, whose density profiles are well described by the Navarro–Frenk–White (NFW) model—a double power law with respect to the halo radius \cite{nfw1,nfw2,nfw3}. In addition to the NFW profile, numerous other models have been proposed based on simulations and astronomical observations \cite{prof1,prof2,prof3,prof4,prof5,prof6,prof7,prof8,prof9}, each describing different radial dependencies of the dark matter halo density. Hong-Sheng Zhao further proposed a generalized double power law density profile characterized by five parameters \cite{zhao}, which encompasses most of the commonly used halo models. 

Dark matter influences not only galactic-scale structures but also the dynamics near the Galactic Center, where its density peaks. It modifies the gravitational field and, consequently, affects the geodesic motion around the central black hole, particularly the black hole shadow and the orbit of the S2 star. Therefore, by using observational data of the black hole shadow and the S2 star orbit, one can place meaningful constraints on the dark matter halo density distribution in the Milky Way.

Previous studies have investigated black holes embedded in various dark matter halos \cite{dmbh1,dmbh2,dmbh3,dmbh4,dmbh5,dmbh6,dmbh7,dmbh8,dmbh8,dmbh9,dmbh10,dmbh11,dmbh12,dmbh13,dmbh14,dmbh15,dmbh16, dmbh17}. However, most of these studies adopt specific forms of dark matter density profiles, which limits their ability to constrain the halo distribution itself. To address this limitation, we adopt Zhao’s generalized double power law profile as the dark matter halo model to explore its influence on the spacetime geometry of the Galactic Center black hole. We aim to constrain the parameters of the dark matter halo by comparing theoretical predictions with observations of the black hole shadow and the S2 star orbit. Assuming that the spacetime of Sgr A$^*$ without dark matter halo is described by the Schwarzschild metric, we first construct the modified black hole metric when it is immersed in Zhao’s dark matter halo. We then derive the null-like and time-like geodesics in this halo-modified spacetime. Based on the resulting geodesic equations, we analyze how the model parameters affect the shadow and orbital motion of the S2 star. Finally, by fitting the model to the observational data, we place constraints on dark matter halo parameters and obtain their posterior distributions.

This paper is organized as follows. In Sec.~\ref{sec2}, we introduce Zhao’s dark matter halo profile and present the formalism to construct the black hole spacetime modified by the dark matter halo. We then investigate how different parameter combinations affect the spacetime metric and its associated geodesic properties. In Sec.~\ref{sec3}, we analyze the null geodesics and the shadow of the dark matter halo modified black hole, examining how the model parameters influence the shadow characteristics. We also use the observed shadow data of Sgr A$^*$ by EHT to constrain the dark matter halo parameters. In Sec.~\ref{sec4}, we investigate the timelike geodesics in the spacetime of the dark matter modified black hole and establish a theoretical framework for analyzing the orbital motion of the S2 star. Using this framework, we perform a Bayesian analysis combined with MCMC sampling to derive the posterior distributions of the model parameters and place quantitative constraints on the dark matter halo density profile. Finally, Sec.~\ref{sec5} summarizes our main results and provides concluding remarks.
Throughout this paper, we use geometrized units with $G = c = 1$. However, for clarity, the fitted parameters in Sec.~\ref{sec4} are presented in their corresponding conventional physical units.

\section{Black Hole Immersed in a Dark Matter Halo}\label{sec2}

The general double power law profile, proposed by Zhao, provides a flexible parameterization for describing the radial density distribution of dark matter halo and astrophysical systems. It is defined as \cite{zhao}
\begin{equation}
\rho(r) = \rho_s \left(\frac{r}{r_s}\right)^{-\gamma} 
\left[\left(\frac{r}{r_s}\right)^{\alpha} + 1\right]^{(\gamma - \beta)/\alpha},
\end{equation}
where $\rho_s$ and $r_s$ denote the characteristic density and scale radius, respectively, while $\alpha$, $\beta$, and $\gamma$ control the shape of the profile. Specifically, $\gamma$ and $\beta$ respectively set the inner and outer slopes, and $\alpha$ determines the sharpness of the transition between these two regimes. All five parameters are non-negative and can be adjusted to model a wide range of density distributions, including cuspy and cored profiles, making the Zhao's model a versatile tool in studies of dark matter halos, galaxies, and other astrophysical systems.

\begin{figure}[htbp]
    \centering
\includegraphics[scale=0.45]{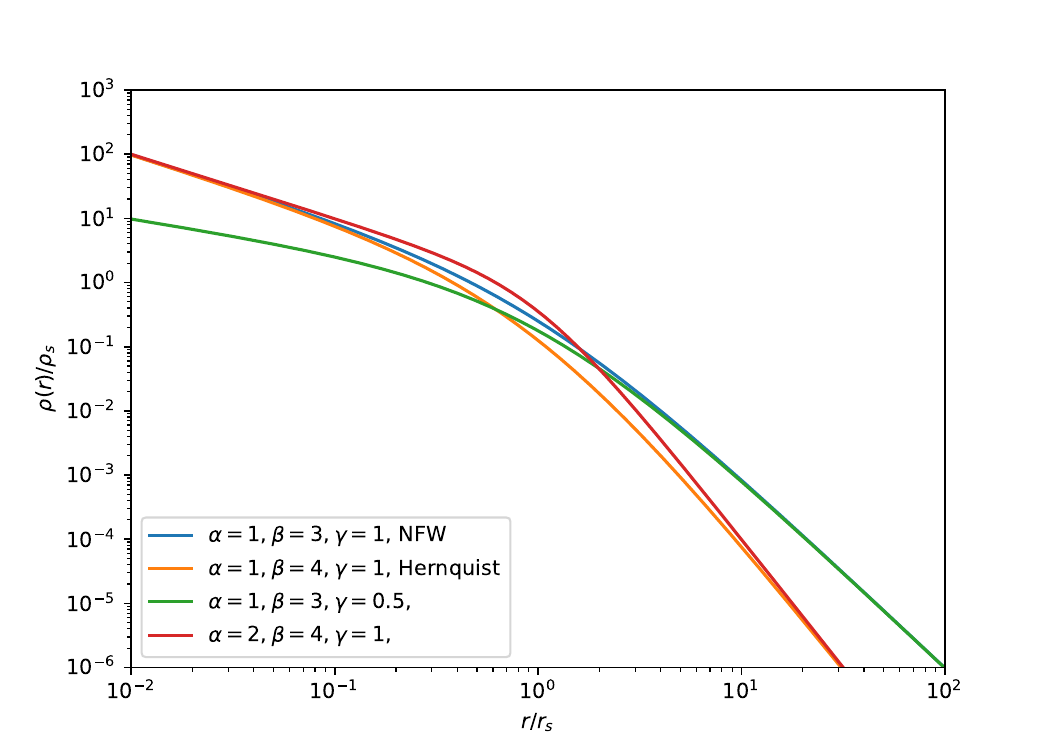}
\caption{Dark matter density profile as a function of the halo radius, for different parameter combinations $(\alpha, \beta, \gamma) = (1, 3, 1), (1, 4, 1), (1, 3, 0.5), (2, 4, 1)$.}
\label{prof}
\end{figure}

To illustrate the effects of the three profile parameters, in Fig.~\ref{prof} we show the normalized dark matter density profiles for different parameter combinations: $(\alpha, \beta, \gamma) = (1, 3, 1), (1, 4, 1), (1, 3, 0.5), (2, 4, 1)$. These four profiles allow a clear comparison of the influence of each parameter. The first and second profiles correspond to the well-known NFW and Hernquist profiles, respectively. It can be seen that the parameter $\alpha$ primarily affects the density around the scale radius, $r \sim r_s$, with larger values of $\alpha$ leading to higher densities in this region. The parameter $\beta$ controls the outer slope of the profile, influencing the density at large radii ($r \gtrsim r_s$); smaller values of $\beta$ result in a more gradual decrease, yielding higher densities at large distances. In contrast, the parameter $\gamma$ governs the inner slope, affecting the density in the central region ($r \lesssim r_s$), where the larger $\gamma$ corresponds to higher central densities. In summary, the three parameters predominantly impact different regions of the halo, relative to the scale radius.

\subsection{Constructing the Dark Matter Halo Modified Black Hole Metric}

The accumulated mass function for Zhao's dark matter halo can be expressed as
\begin{align}\label{md}
M_{D}(r) &= 4 \pi \int_{0}^{r} \rho\left(r^{\prime}\right) r^{\prime 2} d r^{\prime} \nonumber \\
&= \frac{4 \pi \rho_{s} r_{s}^{3}}{\alpha} B\left(\frac{\left(r / r_{s}\right)^{\alpha}}{1+\left(r / r_{s}\right)^{\alpha}}, \frac{3-\gamma}{\alpha}, \frac{\beta-3}{\alpha}\right),
\end{align}
where $B(x, a, b) $ denotes the incomplete beta function, with $ x \in [0, 1] $, $a > 0 $, and $ b > 0 $, which implies that $ \beta > 3 $ and $ \gamma < 3 $ for the function to be well defined. In other words, this expression does not strictly include the NFW profile with $\alpha,\beta,\gamma=(1,3,1)$. However, by introducing small perturbations to this special case, the above formula can still be used to approximately describe the NFW profile as a limiting case.

From this, the tangential velocity profile of a test particle moving in the halo can be defined by
\begin{equation}
v_D^2(r) = \frac{M_D(r)}{r},
\end{equation}
under the Newtonian approximation.

A pure dark matter halo can also be described in the framework of general relativity by a spherically symmetric line element
\begin{equation}
d s^{2}=-F(r) d t^{2}+\frac{d r^{2}}{G(r)}+r^{2}\left(d \theta^{2}+\sin ^{2} \theta d \phi^{2}\right),
\end{equation}
where the functions $F(r)$ and $G(r)$ are the redshift function and the shape function, respectively. In accordance with the Newtonian limit, we assume that the condition $ F(r) = G(r) $ is always satisfied, leading to the relation between the tangential velocity and the metric function \cite{dmbh2},
\begin{equation}
v_{D}^{2}(r) = r \frac{d}{d r} \left(\ln \sqrt{F(r)}\right).
\end{equation}

By integrating the above equation, the metric function $F(r)$ can be obtained explicitly as
\begin{equation}
F(r) = \exp \left(\int_{r}^{\infty} \frac{2 M_{D}(x)}{x^{2}} d x \right),
\end{equation}
which reduces to $F(r) = 1$ in the absence of dark matter. This formalism provides a direct link between the dark matter density distribution and the spacetime geometry.

When the dark matter halo fully determines the energy-momentum tensor, $T_{\mu\nu}^{\rm DM}$, the Einstein field equations take the form:
\begin{equation}
R_{\mu \nu} - \frac{1}{2} R g_{\mu \nu} = \kappa^{2} T_{\mu \nu}^{\rm DM},
\end{equation}
where $\kappa^2 = 8 \pi$. Assuming a diagonal energy-momentum tensor of the form
\begin{equation}
{T}_{\mu}^{\nu} = g^{\nu \alpha} T_{\mu \alpha} = \operatorname{diag}\left[-\rho,\, p_r,\, p,\, p \right],
\end{equation}
this leads to the Einstein equations governing a spherically symmetric dark matter halo,
\begin{align}
\kappa^{2} T_{t}^{t}&=G(r)\left(\frac{1}{r} \frac{G^{\prime}(r)}{G(r)}+\frac{1}{r^{2}}\right)-\frac{1}{r^{2}}, \\
\kappa^{2} T_{r}^{r}&=G(r)\left(\frac{1}{r} \frac{F^{\prime}(r)}{F(r)}+\frac{1}{r^{2}}\right)-\frac{1}{r^{2}}, \\
\kappa^{2} T_{\theta}^{\theta}&=\frac{1}{2} G(r)\left(\frac{F^{\prime \prime}(r) F(r)-F^{\prime 2}(r)}{F^{2}(r)}+\frac{F^{\prime 2}(r)}{2 F^{2}(r)}\right.\nonumber\\
& \quad \left.+\frac{1}{r}\left(\frac{F^{\prime}(r)}{F(r)}+\frac{G^{\prime}(r)}{G(r)}\right)+\frac{F^{\prime}(r) G^{\prime}(r)}{2 F(r) G(r)}\right), \\
\kappa^{2} T_{\phi}^{\phi}&=\kappa^{2} T_{\theta}^{\theta}.
\end{align}
For the combined system of a Schwarzschild black hole and a dark matter halo, the spacetime metric can be expressed as \cite{dmbh2}
\begin{equation}\label{cs}
d s^{2} = -f(r) \, d t^{2} + \frac{1}{g(r)} \, d r^{2} + r^{2} \left( d \theta^{2} + \sin^{2} \theta \, d \phi^{2} \right),
\end{equation}
where
\begin{align}
f(r) &= F(r) + F_1(r), \\
g(r) &= G(r) + F_2(r).
\end{align}
$F_1(r)$ and $F_2(r)$ are the unknown functions that need to be determined by the black hole and dark matter halo parameters. In this case, the Einstein field equations can be written as
\begin{equation}\label{ca}
R_{\mu \nu} - \frac{1}{2} R g_{\mu \nu} = \kappa^{2} \left[ T_{\mu \nu}^{\rm DM} + T_{\mu \nu}^{\rm BH} \right],
\end{equation}
where $T_{\mu \nu}^{\rm BH}$ corresponds to the energy-momentum tensor associated with the matter content of the pure black hole spacetime.

By employing the combined spacetime metric (Eq.~\ref{cs}) and the Einstein field equations (Eq.~\ref{ca}), the resulting expressions are obtained as
\begin{align}\label{fg1}
&\left(G(r)+F_{2}(r)\right)\left(\frac{1}{r^{2}}+\frac{1}{r} \frac{G^{\prime}(r)+F_{2}^{\prime}(r)}{G(r)+F_{2}(r)}\right)\nonumber\\
&=G(r)\left(\frac{1}{r^{2}}+\frac{1}{r} \frac{G^{\prime}(r)}{G(r)}\right), \\
&\left(G(r)+F_{2}(r)\right)\left(\frac{1}{r^{2}}+\frac{1}{r} \frac{F^{\prime}(r)+F_{1}^{\prime}(r)}{F(r)+F_{1}(r)}\right)\nonumber\\
&=G(r)\left(\frac{1}{r^{2}}+\frac{1}{r} \frac{F^{\prime}(r)}{F(r)}\right).\label{fg2}
\end{align}
By imposing the Schwarzschild black hole as a boundary condition, the solutions to the above differential equations can be obtained \cite{dmbh2},
\begin{align}
F_{2}(r)&=-\frac{2 M}{r} ,\\
F_{1}(r)&=\exp \left[\int \frac{G(r)}{G(r)+F_{2}(r)}\left(\frac{1}{r}+\frac{F^{\prime}(r)}{F(r)}\right) -\frac{1}{r} d r\right]\nonumber\\
&\quad -F(r).
\end{align}
As stated above, we have assumed $F(r) = G(r)$, then (Eq.~\ref{fg1}) and (Eq.~\ref{fg2}) implies
\begin{equation}
F_1(r) = F_2(r) = - \frac{2M}{r}.
\end{equation}
Since (Eq.~\ref{cs}) also represents a spherically symmetric metric, and we also assume that $f(r) = g(r)$. Then, we obtain
\begin{equation}\label{fre}
f(r) = g(r) = \exp \left(\int_{r}^{\infty} \frac{2 M_{D}(x)}{x^{2}} d x\right) - \frac{2M}{r},
\end{equation}
which reduces to a Schwarzschild black hole in the absence of dark matter.

The horizon $r_+$ of the dark matter halo modified black hole is determined by
\begin{equation}
f(r) = 0.
\end{equation}

In Table~\ref{cr}, we present the computed horizon radii for black holes embedded in a Zhao's dark matter halo, considering different combinations of the profile parameters $(\alpha, \beta, \gamma) = (1,3,1), (1,4,1), (1,3,0.5), (2,4,1)$. For simplicity, we fix $\rho_s = 0.0001 M^{-2}$ and $r_s = 10 M$. For comparison, the horizon radius of a Schwarzschild black hole ($\rho_s = 0$) is also listed. It is evident that the presence of dark matter increases the horizon radius. Moreover, larger values of $\alpha$ and $\gamma$ and smaller values of $\beta$ correspond to a greater increase in the horizon.

\begin{figure}[htbp]
    \centering
\includegraphics[scale=0.5]{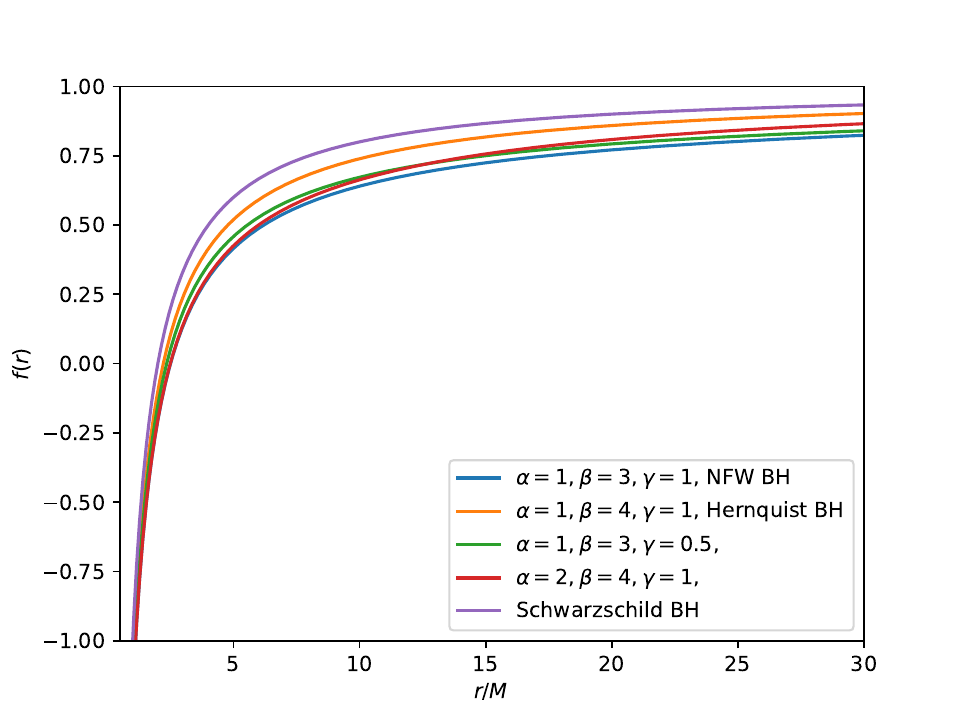}
    \caption{The metric function $f(r)$ is shown as a function of $r/M$ for different combinations of the profile parameters $(\alpha, \beta, \gamma) = (1,3,1), (1,4,1), (1,3,0.5), (2,4,1)$, with $\rho_s = 0.0001M^{-2}$ and $r_s = 10 M$ for simplicity. These parameter sets correspond to different dark matter–modified black hole metrics. For comparison, the Schwarzschild black hole ($\rho_s = 0$) is also shown as a reference.}
    \label{fr1}
\end{figure}

In Fig.~\ref{fr1}, we show the metric function $f(r)$ (Eq.~\ref{fre}) for different combinations of the profile parameters $(\alpha, \beta, \gamma) = (1,3,1), (1,4,1), (1,3,0.5), (2,4,1)$, where we set $\rho_s = 0.0001M^{-2}$ and $r_s = 10 M$ for simplicity. The Schwarzschild black hole is also shown for reference. At small radii, the differences between the black holes are negligible. However, the deviations become more pronounced at larger radii. Across the entire radial range, black holes embedded in a dark matter halo exhibit smaller values of $f(r)$ than the Schwarzschild case, corresponding to a deeper gravitational potential well. This behavior is expected, as the presence of dark matter effectively adds mass or energy to the spacetime. For each case, larger $\alpha$ leads to smaller $f(r)$ and thus a deeper gravitational potential well, consistent with the fact that larger $\alpha$ corresponds to higher dark matter density. The impact of $\alpha$ gradually diminishes at larger radii, since $\alpha$ mainly affects the region where $r/r_s \sim 1$. Similarly, smaller $\beta$ results in a deeper potential well, reflecting higher dark matter density across the large radial range. Larger $\gamma$ also deepens the potential well due to increased inner-halo density; however, the effect of $\gamma$ becomes negligible at large radii, as it predominantly influences the inner region of the halo.

\begin{figure}[htbp]
    \centering
\includegraphics[scale=0.5]{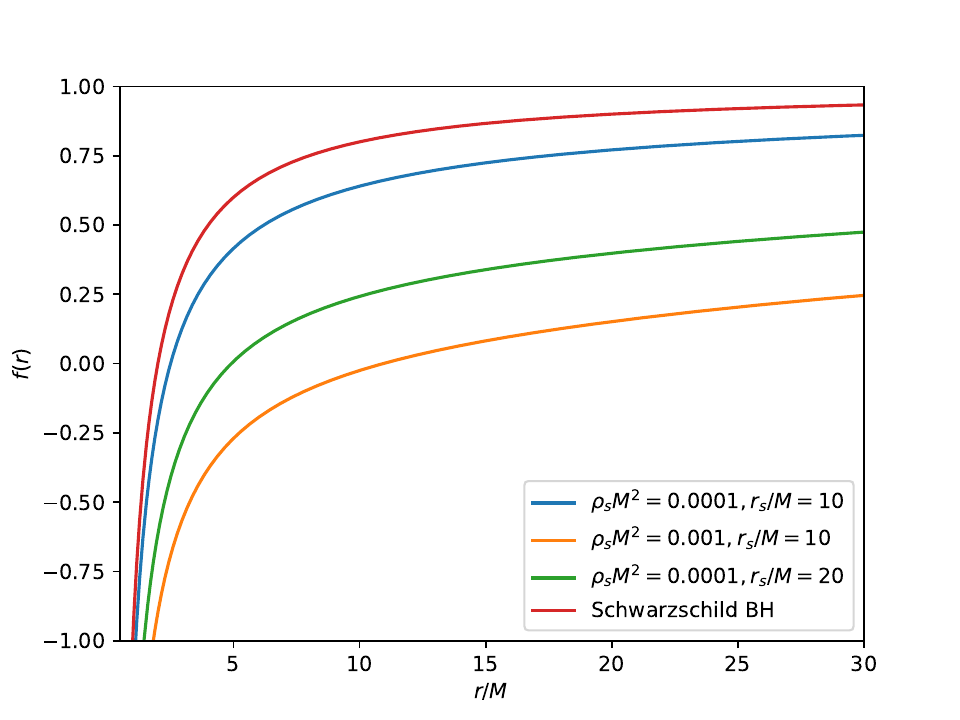}
\caption{The metric function $f(r)$ as a function of $r/M$ for different parameter combinations $(\rho_s*M^{2}, r_s/M) = (0.0001, 10), (0.001, 10), (0.0001, 20)$, with $\alpha, \beta, \gamma = 1, 3, 1$ fixed for simplicity. The Schwarzschild black hole ($\rho_s = 0$) is shown for reference.}
    \label{fr2}
\end{figure}

In Fig.~\ref{fr2}, we show the metric function $f(r)$ as a function of $r/M$ for different parameter combinations $(\rho_s*M^{2}, r_s/M) = (0.0001, 10), (0.001, 10), (0.0001, 20)$, with $\alpha, \beta, \gamma = 1, 3, 1$ fixed. The Schwarzschild black hole, corresponding to $\rho_s = 0$, is also shown for reference. It is evident that a non-zero $\rho_s$ effectively adds mass to the spacetime, resulting in smaller $f(r)$ and a deeper gravitational potential. As expected, larger values of $\rho_s$ lead to smaller $f(r)$ and thus a deeper potential well. Similarly, larger $r_s$, corresponding to a more slowly decreasing dark matter density, also results in smaller $f(r)$.

\subsection{Geodesic Characteristics and Effective Potential}
The Lagrangian that describes the geodesic motion of a particle in a spherically symmetric spacetime can be written as
\begin{equation}
\mathcal{L} = \frac{1}{2} g_{\mu \nu} \dot{x}^{\mu} \dot{x}^{\nu} 
= \frac{1}{2} \left(-f \dot{t}^{2} + \frac{\dot{r}^{2}}{f} + r^{2} \dot{\theta}^{2} + r^{2} \sin^{2} \theta \, \dot{\phi}^{2}\right),
\end{equation}
where $\dot{x}^{\mu} = dx^{\mu}/d\lambda$ is the four-velocity of the particle and $\lambda$ is an affine parameter.  

The spacetime admits two conserved quantities
\begin{equation}
E = -\frac{\partial \mathcal{L}}{\partial \dot{t}} = f \dot{t}, \quad
L = \frac{\partial \mathcal{L}}{\partial \dot{\phi}} = r^{2} \dot{\phi},
\end{equation}
corresponding to the energy and angular momentum of the particle.  

Due to the static and spherically symmetric nature of the metric, the motion can be confined to the equatorial plane, i.e., $\theta = \pi/2$ and $\dot{\theta} = 0$. The radial equation of motion then reads
\begin{equation}
\dot{r}^{2} + V_{\text{eff}}(r) = E^{2}, \quad 
V_{\text{eff}}(r) = f(r) \left(\kappa + \frac{L^{2}}{r^{2}}\right),
\end{equation}
where $V_{\text{eff}}$ is the effective potential and $\kappa = 0,1$ corresponds to null-like and time-like geodesics, respectively.  

For null geodesics ($\kappa = 0$), the photon sphere radius $r_{\rm ph}$ is obtained by solving
\begin{equation}
\frac{dV_{\text{eff}}}{dr} \propto \frac{d}{dr} \left(\frac{f}{r^{2}}\right) = 0.
\end{equation}

For timelike geodesics ($\kappa = 1$), the radius of the innermost stable circular orbit (ISCO) $r_{\rm isco}$ is determined by the conditions
\begin{equation}
\frac{dV_{\text{eff}}}{dr} = 0, \quad \frac{d^{2}V_{\text{eff}}}{dr^{2}} = 0.
\end{equation}

In Table~\ref{cr}, we present the computed characteristic radii, namely the photon sphere and the innermost stable circular orbit (ISCO), for black holes embedded in a Zhao's dark matter halo. We consider different combinations of the profile parameters $(\alpha, \beta, \gamma) = (1,3,1), (1,4,1), (1,3,0.5), (2,4,1)$, with $\rho_s = 0.0001M^{-2}$ and $r_s = 10 M$ fixed for simplicity. For comparison, the Schwarzschild black hole ($\rho_s = 0$) is also included. It is evident that the presence of dark matter increases both characteristic radii. In particular, larger values of $\alpha$ and $\gamma$, and smaller values of $\beta$, correspond to larger photon sphere and ISCO radii.

\begin{table}[]
\begin{tabular}{c@{\hspace{10pt}}c@{\hspace{10pt}}c@{\hspace{10pt}}c@{\hspace{10pt}}c}
\hline
 $\alpha$, $\beta$, $\gamma$   & $r_+/M$   & $r_{ph}/M$     & $r_{isco}/M$ &  \\
\hline
(1, 3, 1)   & 2.503 & 3.771 & 7.098 &  \\
(1, 4, 1)   & 2.217 & 3.336 & 6.426 &  \\
(1, 3, 0.5) & 2.350 & 3.533 & 6.767 &  \\
(2, 4, 1)   & 2.493 & 3.765 & 6.906 &  \\
Schwarzschild  & 2.0     & 3.0       & 6.0     & \\
\hline
\end{tabular}
\caption{Characteristic radii of black holes with a Zhao's dark matter halo, including the horizon, the photon sphere, and the innermost stable circular orbit (ISCO), for different profile parameter combinations $(\alpha, \beta, \gamma) = (1,3,1), (1,4,1), (1,3,0.5), (2,4,1)$, with $\rho_s = 0.0001M^{-2}$ and $r_s = 10 M$ fixed for simplicity. The Schwarzschild black hole ($\rho_s = 0$) is shown for reference.}
\label{cr}
\end{table}

\section{Null Geodesics and Black Hole Shadow}\label{sec3}

For a photon moving in a spherically symmetric spacetime, the impact parameter $b$ is defined as  
\begin{equation}
b = \frac{L}{E},
\end{equation}
where $L$ and $E$ denote the conserved angular momentum and energy of the photon, respectively.  
Using this definition, the null geodesic equation can be expressed in the form 
\begin{equation}\label{fraco}
\left( \frac{dr}{d\phi} \right)^{2} = \frac{r^{4}}{b^{2}} - r^{2} f(r),
\end{equation}
which describes the radial motion of a photon in terms of its trajectory angle $\phi$.  

The critical impact parameter $b_c$ corresponds to the photon orbit at the photon sphere $r_{ph}$, where the gravitational attraction precisely balances the centrifugal repulsion. It is given by  
\begin{equation}
b_c = \frac{r_{ph}}{\sqrt{f(r_{ph})}}.
\end{equation}
Physically, the value of $b$ determines the fate of the photon: for $b < b_c$, the photon is captured by the black hole; for $b = b_c$, it undergoes an unstable circular motion at the photon sphere; and for $b > b_c$, the photon is deflected by the gravitational field and eventually escapes to infinity.  

The degree of light ray deflection can be characterized by the total number of orbits
\begin{equation}
n = \frac{\phi}{2\pi},
\end{equation}
which quantifies how many times the light rays wind around the black hole before escaping. This quantity encapsulates the gravitational lensing behavior of photons near compact objects and serves as an important quantity in black hole shadow and strong lensing analyzes.

\begin{figure}[htbp]
    \centering
\includegraphics[scale=0.5]{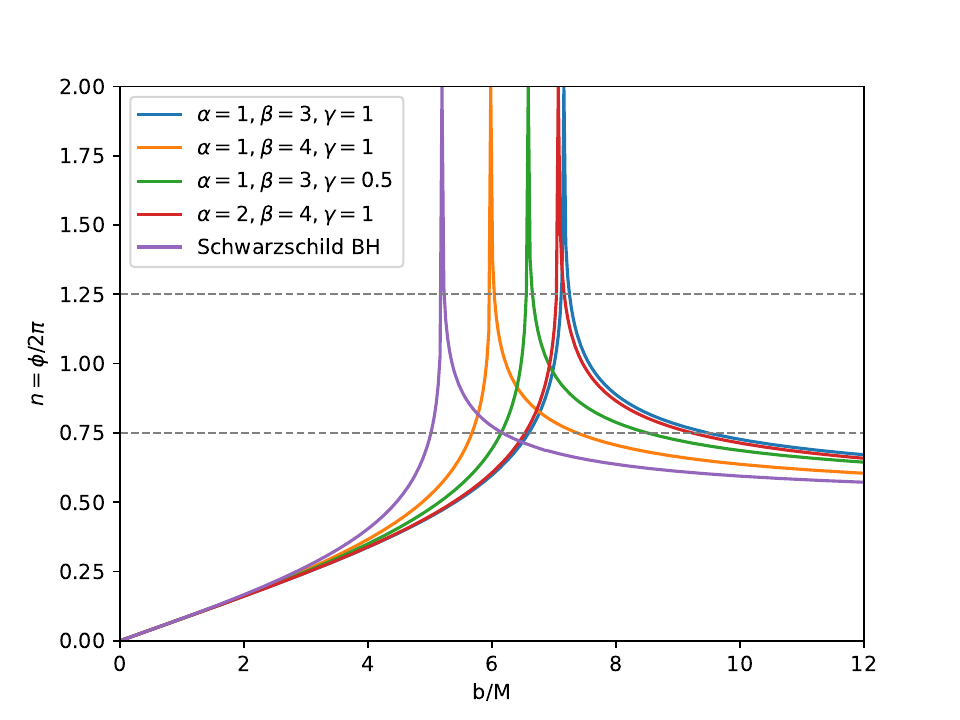}
\caption{Total number of orbits $n = \phi / 2\pi$ as a function of the impact parameter $b$, obtained for different combinations of the profile parameters $(\alpha, \beta, \gamma) = (1, 3, 1)$, $(1, 4, 1)$, $(1, 3, 0.5)$, and $(2, 4, 1)$. For simplicity, the parameters are fixed as $\rho_s = 0.0001M^{-2}$ and $r_s = 10M$. The corresponding result for the Schwarzschild black hole ($\rho_s = 0$) is also shown for comparison.}
    \label{phib}
\end{figure}

In Fig.~\ref{phib}, we present the total number of orbits $n = \phi / 2\pi$ as a function of the impact parameter $b$, obtained by numerically solving (Eq.~\ref{fraco}) for different combinations of the profile parameters $(\alpha, \beta, \gamma) = (1, 3, 1)$, $(1, 4, 1)$, $(1, 3, 0.5)$, and $(2, 4, 1)$. For simplicity, we set $\rho_s = 0.0001M^{-2}$ and $r_s = 10M$. The corresponding result for the Schwarzschild black hole ($\rho_s = 0$) is also shown for comparison.  It is evident that the presence of a dark matter halo leads to an overall shift of the peak positions toward larger impact parameters. For smaller impact parameters ($b \lesssim 5$), larger values of $\alpha$ and $\beta$ and smaller values of $\gamma$ result in a greater number of orbits $n$. In contrast, for larger impact parameters ($b \gtrsim 8$), the same parameter trends produce smaller values of $n$. These behaviors reflect the influence of different dark matter halo profiles on the photon trajectories and consequently on the gravitational lensing characteristics near the black hole.

\begin{figure}[htbp]
    \centering
\includegraphics[scale=0.45]{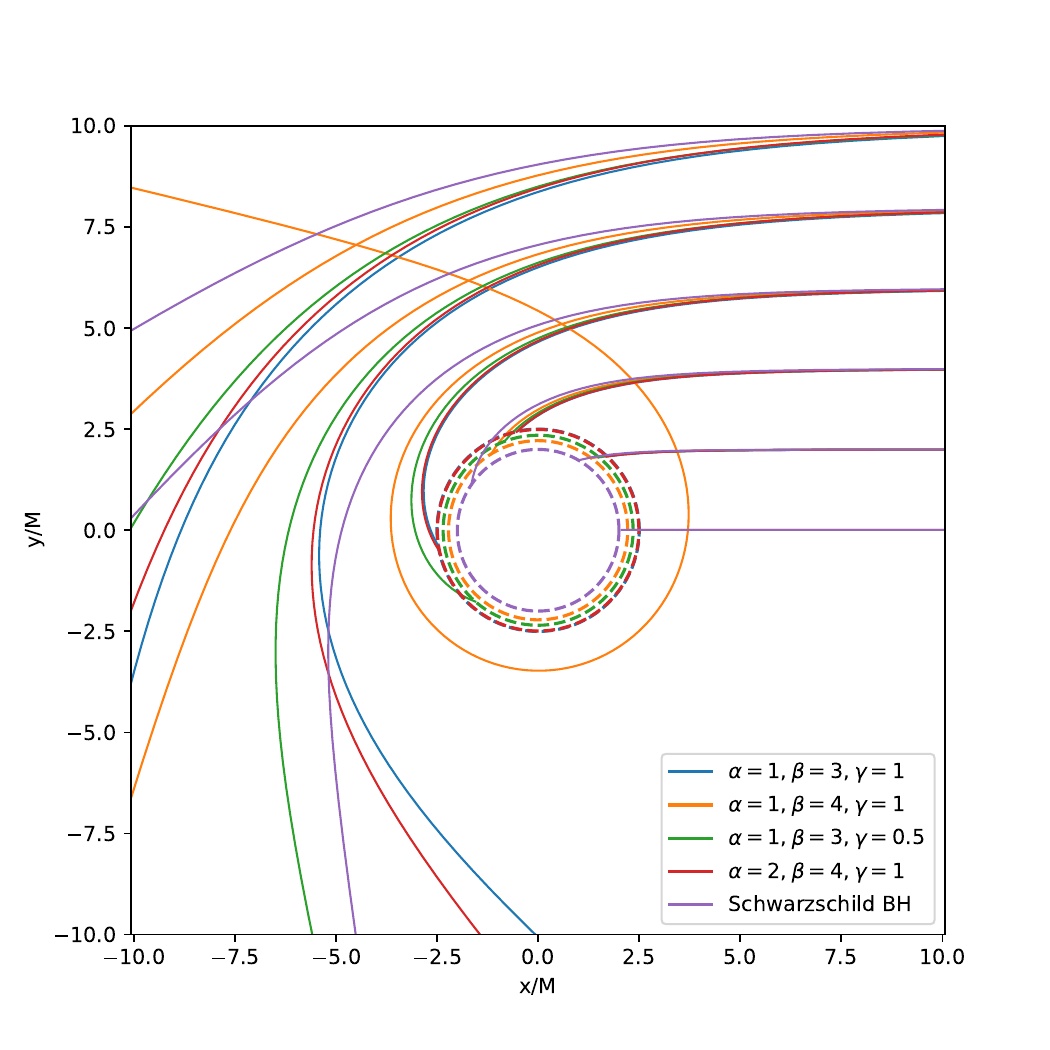}
\caption{Photon trajectories for impact parameters $b = 0, 2, 4, 6, 8, 10$ in Euclidean coordinates $x = r \cos\phi$ and $y = r \sin\phi$, obtained for different combinations of the profile parameters $(\alpha, \beta, \gamma) = (1, 3, 1)$, $(1, 4, 1)$, $(1, 3, 0.5)$, and $(2, 4, 1)$. For simplicity, we set $\rho_s = 0.0001M^{-2}$ and $r_s = 10M$. The corresponding result for the Schwarzschild black hole ($\rho_s = 0$) is also shown for comparison. Solid lines represent photon trajectories, while dashed lines indicate the horizons.}
    \label{rays}
\end{figure}

In Fig.~\ref{rays}, we present the photon trajectories for impact parameters $b = 0, 2, 4, 6, 8, 10$ in Euclidean coordinates, defined as $x = r \cos\phi$ and $y = r \sin\phi$, for different combinations of the profile parameters $(\alpha, \beta, \gamma) = (1, 3, 1)$, $(1, 4, 1)$, $(1, 3, 0.5)$, and $(2, 4, 1)$. For simplicity, we adopt $\rho_s = 0.0001M^{-2}$ and $r_s = 10M$. The trajectory corresponding to the Schwarzschild black hole ($\rho_s = 0$) is also plotted for comparison.  It can be seen that a non-zero $\rho_s$, representing the presence of a dark matter halo, enhances the deflection of photon trajectories, indicating a stronger effective gravitational field. Moreover, larger values of $\alpha$ and $\gamma$, and smaller values of $\beta$, lead to more pronounced light bending and an increase in the apparent horizon radius.

In addition, for a static observer located at a radial distance $r_o$, the radius of the black hole shadow $R_s$ is given by \cite{rs}
\begin{equation}
R_s = b_c \sqrt{f(r_o)} = r_{ph} \sqrt{\frac{f(r_o)}{f(r_{ph})}}.
\end{equation}
In the limit $r_o \rightarrow \infty$, this expression reduces to $R_s = b_c$. 

\begin{figure}[htbp]
    \centering
\includegraphics[scale=0.53]{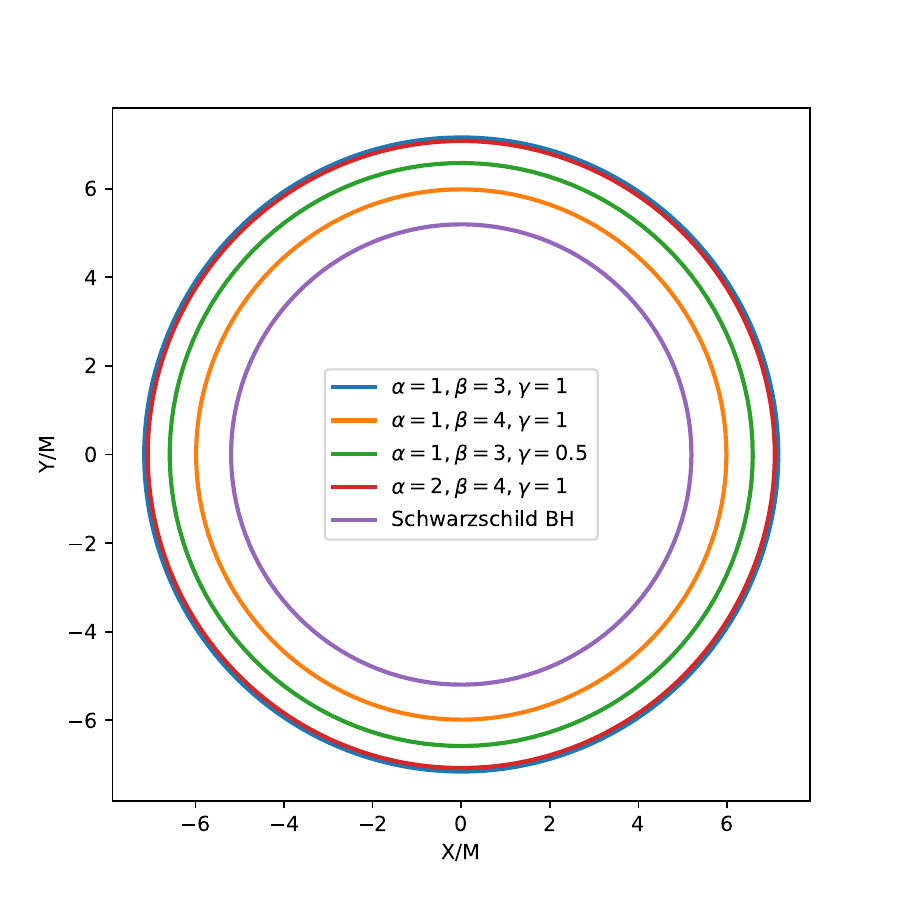}
\caption{Black hole shadow radius projected onto the observer’s plane, where $X$ and $Y$ denote the celestial coordinates of the black hole image. Results are shown for different combinations of the profile parameters $(\alpha, \beta, \gamma) = (1, 3, 1)$, $(1, 4, 1)$, $(1, 3, 0.5)$, and $(2, 4, 1)$. For simplicity, we set $\rho_s = 0.0001M^{-2}$ and $r_s = 10M$. The corresponding shadow of the Schwarzschild black hole ($\rho_s = 0$) is also displayed for comparison.}
    \label{shd}
\end{figure}

In Fig.~\ref{shd}, we show the black hole shadow radius projected onto the observer’s plane, where $X$ and $Y$ denote the celestial coordinates of the black hole image. The results are presented for different combinations of the profile parameters $(\alpha, \beta, \gamma) = (1, 3, 1)$, $(1, 4, 1)$, $(1, 3, 0.5)$, and $(2, 4, 1)$. For simplicity, we adopt $\rho_s = 0.0001M^{-2}$ and $r_s = 10M$. The corresponding shadow of the Schwarzschild black hole ($\rho_s = 0$) is also plotted for comparison.  It is evident that the presence of dark matter enlarges the apparent shadow radius. Furthermore, larger values of $\alpha$ and $\gamma$, together with smaller values of $\beta$, lead to an overall increase in the size of the black hole shadow.

\begin{figure*}[htbp]
    \centering
\includegraphics[scale=0.5]{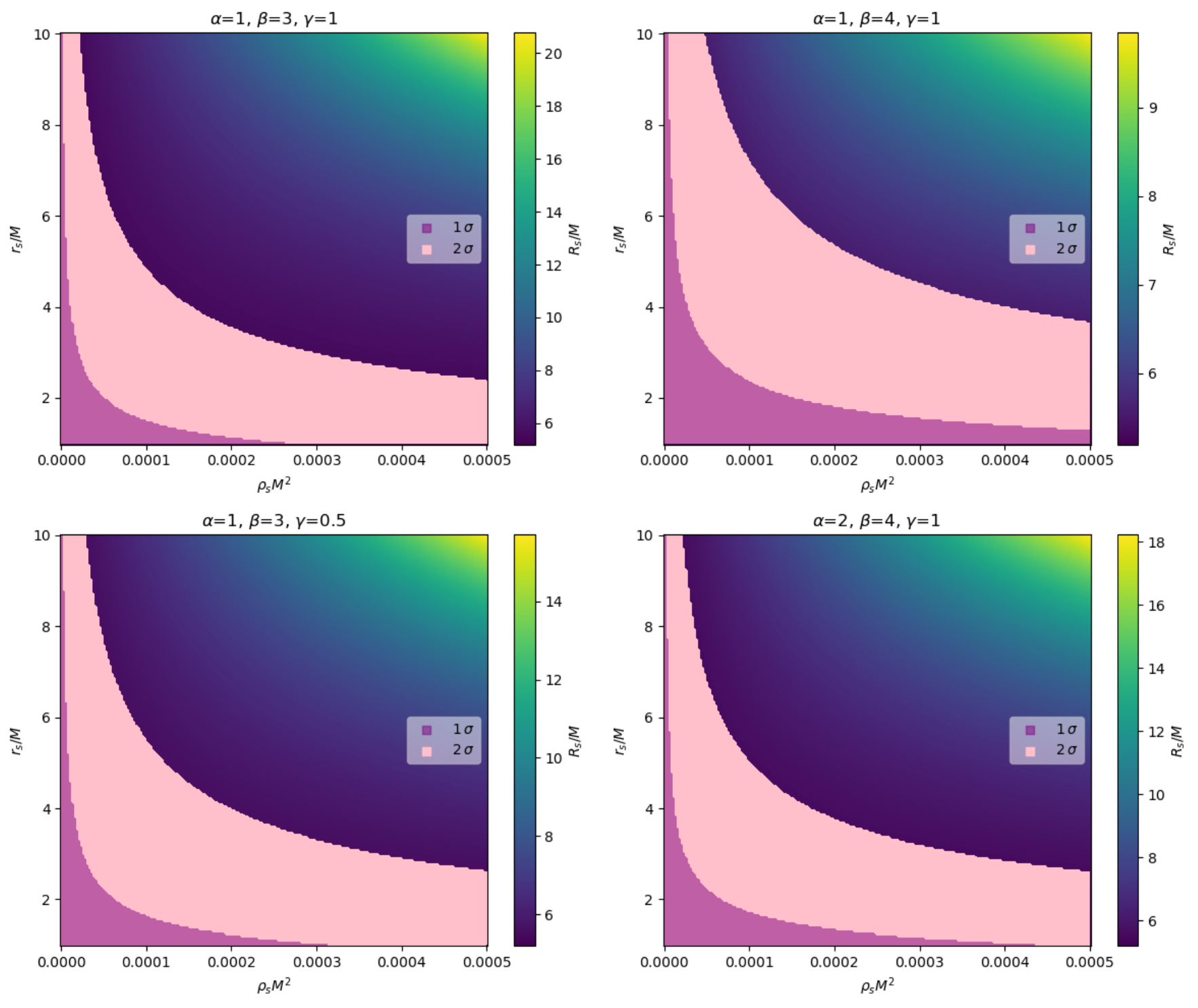}
\caption{Two-dimensional parameter space and observational constraints of Sgr A* shadows as functions of $\rho_s M^2$ and $r_s/M$, for different profile parameter combinations $(\alpha, \beta, \gamma) = (1, 3, 1), (1, 4, 1), (1, 3, 0.5), (2, 4, 1)$. The color bar on the right of each subplot indicates the shadow radius. The pink and purple regions correspond to the $2\sigma$ and $1\sigma$ confidence regions of the Sgr A* shadow observations, respectively. The vertical axis, where $\rho_s = 0$, corresponds to the Schwarzschild black hole.}
    \label{para}
\end{figure*}

In Fig.~\ref{para}, we present the two-dimensional parameter space and the constraints using the Sgr A* shadow data observed by EHT, as functions of $\rho_s M^2$ and $r_s/M$, for different profile parameter combinations $(\alpha, \beta, \gamma) = (1, 3, 1), (1, 4, 1), (1, 3, 0.5), (2, 4, 1)$. The color bar on the right of each subplot indicates the shadow radius. The pink and purple regions correspond to the $2\sigma$ and $1\sigma$ confidence regions of the Sgr A* shadow observations, respectively. The vertical axis, where $\rho_s = 0$, corresponds to the Schwarzschild black hole. It is evident that larger values of $\beta$ and smaller values of $\alpha$ and $\gamma$ result in a wider parameter space consistent with the Sgr A* shadow observations. Moreover, $\rho_s M^2$ and $r_s/M$ exhibit a negative correlation within the allowed parameter space.

\section{Timelike Geodesics and S2 Star Motions}\label{sec4}

The second-order geodesic equations for massive particle objects like S2 star take the form of
\begin{equation}
\frac{d^{2} x^{\alpha}}{d \lambda^{2}} + \Gamma_{\beta \rho}^{\alpha} \frac{d x^{\beta}}{d \lambda} \frac{d x^{\rho}}{d \lambda} = 0,
\end{equation}
where $\lambda$ is an affine parameter along the particle's worldline and $\Gamma^{\mu}_{\nu\sigma}$ are the Christoffel symbols associated with the spacetime metric.  

Since the metric is static and spherically symmetric, we can choose the orbital plane of S2 star to be the equatorial plane and restrict the S2 star motion within it, i.e., $\theta = \pi/2$ and $\dot{\theta} = 0$. Then, explicitly, the geodesic equations reduce to
\begin{align}
\ddot{t} &= -2 \, \Gamma^t_{tr} \, \dot{t} \, \dot{r}, \\
\ddot{r} &= - \Gamma^r_{tt} \, \dot{t}^2 - \Gamma^r_{rr} \, \dot{r}^2 - \Gamma^r_{\phi\phi} \, \dot{\phi}^2,\\
\ddot{\theta} &= 0,\\
\ddot{\phi} &= -2 \, \Gamma^\phi_{r\phi} \, \dot{r} \, \dot{\phi},
\end{align}
where a dot denotes differentiation with respect to $\lambda$.  To solve these second-order differential equations, one must specify the initial conditions. Since $\theta$ is fixed, it does not require an initial condition. For simplicity, we assume $\lambda = 0$ corresponds to the time of apocentre passage, yielding
\begin{align}
t(0) &= t_p - T/2, \\
r(0) &= a(1+e) \equiv r_0,\\
\dot{r}(0) &= 0, \\
\phi(0) &= \pi,\\
\dot{\phi}(0) &= L/r_0^2 \equiv \dot\phi_0, 
\end{align}
where $t_p$ is the time of pericenter passage, and we adopt the specific angular momentum $L = \sqrt{(1 - e^2)aM} $ of a Keplerian orbit, since the motion can be well approximated by classical Newtonian gravity near the apocenter passage. $T$ is the orbital period, $e$ is the orbital eccentricity, and $a$ is the semi-major axis of the orbit. The initial value of $\dot{t}$ can be obtained from the normalization condition of the four-velocity, $g_{\mu\nu}\dot{x}^\mu \dot{x}^\nu = -1$, giving
\begin{equation}
\dot{t}(0)= \sqrt{\frac{1 + r_0^2 \dot{\phi}_0^2}{f(r_0)}}.
\end{equation}

The solutions $t(\lambda)$, $r(\lambda)$, and $\phi(\lambda)$ provide the S2 star's trajectory in the orbital plane with polar coordinates. The radial velocity in the orbital plane is
\begin{equation}
v_r = \frac{dr}{dt} = \frac{\dot{r}}{\dot{t}}.
\end{equation}
The particle's position and velocity expressed in Cartesian coordinates in the orbital plane, are
\begin{align}
x_{\rm orb} &= r \cos\phi,\\
y_{\rm orb} &= r \sin\phi,\\
v_{x,{\rm orb}} &= v_r \cos\phi,\\
v_{y,{\rm orb}} &= v_r \sin\phi.
\end{align}

In order to compare the orbit with observations, we need to project the trajectory of S2 star onto the observer's plane using the Thiele-Innes elements \cite{angl}:
\begin{align}
A &= \cos\Omega \cos\omega - \sin\Omega \sin\omega \cos i, \\
B &= \sin\Omega \cos\omega + \cos\Omega \sin\omega \cos i, \\
C &= -\sin\omega \sin i, \\
F &= -\cos\Omega \sin\omega - \sin\Omega \cos\omega \cos i, \\
G &= -\sin\Omega \sin\omega + \cos\Omega \cos\omega \cos i, \\
H &= -\cos\omega \sin i,
\end{align}
where $\Omega$ is the longitude of the ascending node, $\omega$ is the argument of the pericentre, and $i$ is the orbital inclination.  

The positions and velocities of the S2 stat in the observer plane are then given by \cite{angl}
\begin{align}
x_{\rm obs} &= A x_{\rm orb} + F y_{\rm orb},\\
y_{\rm obs} &= B x_{\rm orb} + G y_{\rm orb},\\
z_{\rm obs} &= -(C x_{\rm orb} + H y_{\rm orb}),\\
v_{x,{\rm obs}} &= A v_{x,{\rm orb}} + F v_{y,{\rm orb}},\\
v_{y,{\rm obs}} &= B v_{x,{\rm orb}} + G v_{y,{\rm orb}},\\
v_{z,{\rm obs}} &= -(C v_{x,{\rm orb}} + H v_{y,{\rm orb}}).
\end{align}

Before applying the comparison with observational data, it is necessary to account for both classical and relativistic effects that can modify the observed orbit of S2 star. The observed position and radial velocity should account for the zero-point offsets and drifts of Sgr A$^*$. because all measurements are referenced to it, which could be represented by five parameters $(x_0, y_0, v_{x,0}, v_{y,0}, v_{z,0})$. In addition, we also need to consider the changes in light travel time and relativistic effects caused by S2’s orbital motion. In particular, the R{\o}mer delay relates the observed time $t_o$ to the emission time $t_e$ via \cite{crec1}
\begin{equation}
t_e \approx  t_o - z_{\rm obs}(1-v_{z,{\rm obs}} ),
\end{equation}
and the observed radial velocity is affected by both the Doppler effect and the gravitational redshift, which can be approximated as \cite{crec2}
\begin{equation}
v_{z,{\rm crt}} \approx  v + \frac{M}{r},
\end{equation}
where $v = \sqrt{v_{x,{\rm obs}}^2 + v_{y,{\rm obs}}^2 + v_{z,{\rm obs}}^2}$ is the total observed velocity of the particle.

Taking all these effects into account, the corrected observed positions in the celestial coordinates and radial velocity of S2 star are
\begin{align}
x_{\rm obs}(t_o) &= x_{\rm obs}(t_e) + x_0 + v_{x,0} \, (t_o - 2009),\\
y_{\rm obs}(t_o) &= y_{\rm obs}(t_e) + y_0 + v_{y,0} \, (t_o - 2009),\\
v_{z,{\rm obs}}(t_o) &= v_{z,{\rm obs}}(t_e) + v_{z,0} + v_{z,{\rm crt}}.
\end{align}
where the year 2009 is chosen as the reference epoch for the position and proper motion of Sgr A$^*$.

In this framework, the trajectory of the S2 star is completely characterized by specifying the following set of 19 parameters:
\begin{itemize}
    \item orbital parameter: $M, \,R, \,T, \,t_p, \,a, \,e, \,i, \,\Omega, \,\omega$

    \item offsets
and drifts parameters: $\,x_0, \,y_0, \,v_{x0}, \,v_{y0}, \,v_{z0}$

    \item dark matter halo parameters: $\,\rho_s, \,r_s, \,\alpha, \,\beta, \,\gamma$ 
\end{itemize}

The model is now ready to be compared with the observational data.

\subsection{Data and Bayesian Inference Methodology}

We use the publicly available astrometric and spectroscopic data of S2 as follows:

\begin{itemize}
    \item A total of 145 astrometric measurements of the S2 star position were extract from Table~5 of \cite{pdata1}, covering the observational period between 1992.225 and 2016.38. These observations originate from multiple facilities and instruments:  
    (a) data obtained between 1992.224 and 2002 were recorded using the SHARP speckle camera mounted on the ESO New Technology Telescope (NTT; \cite{pdata2}), with typical positional uncertainties of approximately 3.8~mas;  
    (b) measurements from 2002 to 2016.38 were taken with the NAOS-CONICA (NACO) adaptive optics infrared camera on the Very Large Telescope (VLT; \cite{pdata3,pdata4}), achieving a root-mean-square astrometric precision of about 400~$\mu$as.
    
    \item In addition, 44 spectroscopic observations of the Brackett-$\gamma$ emission line were used to determine the radial velocity of S2 \cite{pdata1}. Spectra obtained prior to 2003 were measured with the NIRC2 adaptive optics imager and spectrograph at the Keck Observatory \cite{ss1}, while those taken after 2002 were acquired with the SINFONI instrument, an adaptive optics–assisted integral field spectrograph on the VLT \cite{vdata1,vdata2}.
\end{itemize}

To explore and constrain the parameter space, we use the observational data described above and the orbital model of S2 star outlined in the previous section. Specifically, we adopt a Bayesian inference approach and employ the \texttt{emcee} package \cite{emcee}, which implements a Markov Chain Monte Carlo (MCMC) algorithm to sample the posterior distributions of the model parameters.

The posterior probability distribution $ p(\Theta|\text{D}) $ represents the updated probability of the model parameters $ \Theta $ after taking into account the observed data $ \text{D} $. It can be expressed, up to a normalization constant, as  
\begin{equation}
\log p(\Theta|\text{D}) \propto \log \mathcal{L}(\text{D}|\Theta) + \log q(\Theta),
\end{equation}
where $ \mathcal{L}(\text{D}|\Theta) $ represents the likelihood function that quantifies the probability of observing the data $\text{D} $ given a set of model parameters $ \Theta $, and $ q(\Theta)$ denotes the prior distribution that encodes our prior knowledge or assumptions about $\Theta $ before considering the data. 

The likelihood function for a given set of model parameters is evaluated by comparing the model predictions with the observational data as
\begin{equation}
\log \mathcal{L}(\text{D}|\Theta) = \log \mathcal{L}_P(\text{D}|\Theta) + \log \mathcal{L}_V(\text{D}|\Theta),
\end{equation}
where $\log \mathcal{L}_P$ represents the contribution from the astrometric (sky-projected) positions of S2, and $\log \mathcal{L}_V$ corresponds to the spectroscopic radial velocity measurements. These contributions are explicitly given by
\begin{align}
\log \mathcal{L}_{P} &= -\sum_{i} \frac{\left(x_{\rm data}^{i} - x_{\rm obs}^{i}\right)^{2}}{2 \left(\sigma_{x,{\rm data}}^{i}\right)^{2}}
                        - \sum_{i} \frac{\left(y_{\rm data}^{i} - y_{\rm obs}^{i}\right)^{2}}{2 \left(\sigma_{y,{\rm data}}^{i}\right)^{2}},\\
\log \mathcal{L}_{V} &= -\sum_{i} \frac{\left(v_{z,{\rm data}}^{i} - v_{z,{\rm obs}}^{i}\right)^{2}}{2 \left(\sigma_{v_z,{\rm data}}^{i}\right)^{2}},
\end{align}
where $x_{\rm data}^{i}$ and $y_{\rm data}^{i}$ are the observed astrometric positions of the $i$-th data point on the sky, and $\sigma_{x,{\rm data}}^{i}$ and $\sigma_{y,{\rm data}}^{i}$ denote the corresponding measurement uncertainties. Similarly, $v_{z,{\rm data}}^{i}$ is the observed radial velocity of the $i$-th data point, with uncertainty $\sigma_{v_z,{\rm data}}^{i}$. $x_{\rm obs}^{i}$, $y_{\rm obs}^{i}$ and $v_{z, \rm obs}^{i}$ denote the model-predicted values of the corresponding observables for the $i$-th data point, computed from a given set of model parameters.

We adopt the same Gaussian priors for the orbital parameters as in \cite{prior1} and for the zero-point offset and drift parameters as in \cite{prior2}, reflecting that these quantities are already reasonably well constrained by previous studies. For the dark matter halo parameters, we apply Gaussian priors to $\rho_s$ and $r_s$, informed by previous fits of a Navarro-Frenk-White (NFW) dark matter halo to the Milky Way rotation curve \cite{prior3}. In contrast, for the shape parameters $\alpha$, $\beta$, and $\gamma$, we employ bounded uniform priors due to the lack of existing constraints, as exploring their plausible ranges is one of the primary goals of this study.  

The adopted priors are summarized in Table~\ref{pri}. The parameter ranges for the dark matter halo in our analysis are chosen to cover a sufficiently broad and physically meaningful region. Specifically, we set $3 \le \beta \le 10$ to ensure that (Eq.~\ref{md}) remains valid (for case $\beta = 3$, we apply a small positive perturbation of order $~10^{-6}$) and the halo density decreases rapidly enough at large radii to yield a finite total mass, but not excessively fast. The upper limit $\gamma \le 3$ is imposed in order to keep (Eq.~\ref{md}) valid (for case $\gamma = 3$, we apply a small negative perturbation of order $~10^{-6}$) and to prevent an overly steep central cusp, consistent with numerical simulations \cite{dms1,dms2}. Finally, we set $\alpha \le 5$ to guarantee a smooth transition of the density profile between the inner and outer regions of the dark matter halo.


\begin{table}[h]
\centering
\begin{tabular}{l c c c c}
\hline
Parameter & $\mu$ & $\sigma$ & Reference & Uniform \\
\hline
$M~(10^6 M_{\odot})$ & 4.261 & 0.012 & G & -- \\
$R~(\mathrm{kpc})$ & 8.2467 & 0.0093 & G & -- \\
\hline
$T~(\mathrm{yr})$ & 16.0455 & 0.0013 & G & -- \\
$t_p~(\mathrm{yr})$ & 2018.37800 & 0.00017 & G & -- \\
$a~(\mathrm{mas})$ & 125.058 & 0.044 & G & -- \\
$e$ & 0.884649 & 0.000079 & G & -- \\
$i~(^{\circ})$ & 134.567 & 0.033 & G & -- \\
$\omega~(^{\circ})$ & 66.263 & 0.030 & G & -- \\
$\Omega~(^{\circ})$ & 228.171 & 0.031 & G & -- \\
\hline
$x_0~(\mathrm{mas})$ & -0.2 & 0.2 & P & -- \\
$y_0~(\mathrm{mas})$ & 0.1 & 0.2 & P & -- \\
$v_{x0}~(\mathrm{mas/yr})$ & 0.05 & 0.1 & P & -- \\
$v_{y0}~(\mathrm{mas/yr})$ & 0.06 & 0.1 & P & -- \\
$v_{z0}$ & -1.6 & 1.4 & G & -- \\
\hline
$\alpha$ & -- & -- & -- & $[0, \, 5]$ \\
$\beta$ & -- & -- & -- & $[3, \, 10]$ \\
$\gamma$ & -- & -- & -- & $[0, 3]$ \\
$\rho_s~(M_\odot/\mathrm{pc}^3)$ & 0.00718 & 0.0009 & N & -- \\
$r_s~(\mathrm{kpc})$ & 15.3 & 1.1 & N & -- \\
\hline
\end{tabular}
\caption{Summary of the priors used in our MCMC analysis. For the orbital parameters of S2, we adopt the most recent measurements from \cite{prior1} (G), and the priors for the reference-frame offsets are taken from \cite{prior2} (P), as well as the reference of NFW model for $\rho_s$ and $r_s$ \cite{prior3} (N). These are implemented as Gaussian priors and are presented in the second and third columns, with the corresponding references in the fourth column. The dark matter halo parameters $\alpha$, $\beta$ and $\gamma$ are assigned uniform priors with estimated ranges as listed in the last column.}
\label{pri}
\end{table}

Subsequently, we perform MCMC analysis with $4*19 =76$ walkers, each taking 6000 steps (including an initial burn-in phase of 2000 steps), to rigorously sample the posterior distribution defined by the aforementioned log-likelihood function and prior distributions. This approach enables an efficient exploration of the multi-dimensional parameter space, allowing us to estimate the most probable values of the model parameters and to quantify their corresponding uncertainties.

\subsection{MCMC Results and Orbital Fitting}

\begin{figure*}[htbp]
    \centering
\includegraphics[scale=0.17]{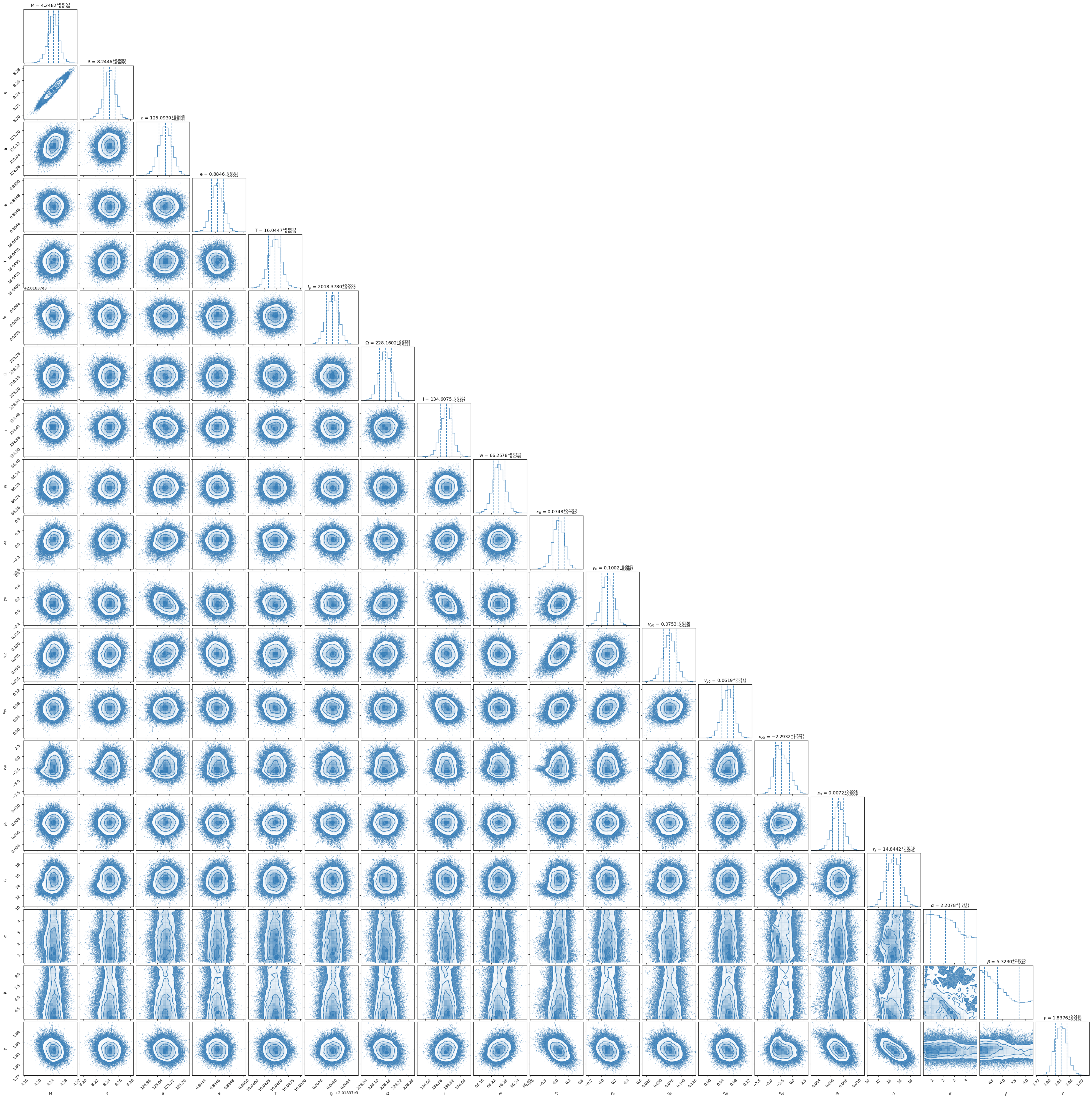}
\caption{Posterior distributions of the MCMC samples for the parameters of the S2 star orbital model. In each subplot, The contour levels in the 2D marginalized distributions correspond to the $1\sigma$, $2\sigma$, and $3\sigma$ credible regions in two dimensions, enclosing approximately 39\%, 86\%, and 99\% of the posterior probability, respectively. The histograms display the marginal posterior distributions of each parameter, with vertical lines indicating the median values and the corresponding $1\sigma$ credible intervals, which are also annotated above each histogram.}
    \label{bays}
\end{figure*}

In Fig.~\ref{bays}, we present the posterior distributions of the MCMC samples for the parameters of the S2 star orbital model, derived from the S2 observational data. The contours enclosing approximately 39\%, 86\%, and 99\% of the 2 dimensional posterior probability are shown, while the histograms display the marginalized posterior distributions for each parameter, with the median values and $1\sigma$ credible intervals indicated. The figure also illustrates the correlations between parameter pairs. For instance, $M$ and $R$ exhibit a positive correlation, indicating that simultaneous increases in $M$ and $R$ improve the agreement between the model and the observational data.

In Table~\ref{fit}, we present the best-fitting (median) values and the corresponding $1\sigma$ credible intervals of the MCMC samples for the parameters of the S2 orbital model. For comparison, we also list the best-fitting values from \cite{pdata1} and \cite{prior1}, which were obtained using Keplerian and post-Newtonian approaches, respectively, assuming the absence of a dark matter halo. In contrast, our analysis solves the full geodesic equations with dark matter halo.  The posterior distributions of the parameters in these reference works are approximately Gaussian, yielding symmetric $1\sigma$ intervals, whereas our results exhibit slight deviations from Gaussianity, resulting in mildly asymmetric $1\sigma$ bounds for some parameters. 

From Table~\ref{fit}, we can see that the best-fitting (median) values of the main orbital parameters ($M, R, T, t_p, a, e, i, \Omega, \omega$) are broadly consistent with, yet slightly different from, those reported in previous studies, indicating that the dark matter density near the Galactic Center is low but may exert a small, non-negligible influence on the S2 orbit. Significant deviations are observed in the zero-point offset and drift parameters $(x_0, y_0, v_{x0}, v_{y0}, v_{z0})$, which primarily reflect the current observational uncertainties. The dark matter halo parameters $\rho_s$, $r_s$, and $\gamma$ are better constrained than the other two parameters, $\alpha$ and $\beta$. This can also be seen in Fig.~\ref{bays}, where the posterior distributions related to $\alpha$ and $\beta$ appear widely dispersed and poorly constrained, indicating significant degeneracy and uncertainty in these parameters. This is likely because the S2 orbit lies well within the inner region of the dark matter halo, where only the parameters governing the inner density profile ($\gamma$) have a significant effect. Nevertheless, we have successfully obtained meaningful constraints on the dark matter halo parameters of the Milky Way.

Using the best-fitting results of parameters, we can estimate the local dark matter density in the Solar System to be: 
\begin{align}
\rho_\odot &= 0.0145~M_\odot/\mathrm{pc}^3 \nonumber\\
&= 9.8085 * 10^{-22}~\mathrm{kg/m^3}\nonumber\\ 
&= 0.5503~\mathrm{GeV/cm^3} .  
\end{align}
This is well consistent with previous studies using different observation approach \cite{prior3,prior4,prior5,prior6}. We also compute the total dark matter mass within the S2 orbital region, approximated as a sphere with radius equal to the major axis of the S2 orbit ($\sim 230 \,mas$), which yields $\lesssim 0.3642\% ~M$ of the central black hole mass, consistent in order of magnitude with previous studies \cite{mpro}. This indicates that the dark matter has only a tiny effect on the dynamics near our Galactic Center.  
Furthermore, we can estimate the shadow radius of Sgr~A* using the best-fitting dark matter halo parameters. We obtain
\begin{equation}
R_s \approx 5.19615547 ~M.
\end{equation}
The shadow radius agrees with that of the Schwarzschild black hole ($5.19615242~M$) up to the sixth decimal place, which lies within the $1\sigma$ region of the Sgr~A* shadow observation, as illustrated in Fig.~\ref{para}.

\begin{table*}[]
\centering
\renewcommand{\arraystretch}{1.3}
\setlength{\tabcolsep}{4pt}
\begin{tabular}{l|c|c|c}
\hline
 & Gillessen et al. (2017) & Gravity Collab. et al. (2020) & {This work} \\
Parameter \textbackslash\,  Model & Keplerian & Post Newtonian & Geodesic\\
\hline
$M_\bullet \,(10^6 M_\odot)$ & $4.3 \pm 0.15$ & $4.261 \pm 0.012$ & $4.2482^{+0.0153}_{-0.0158}$ \\
$R_\bullet$ (kpc) & $8.17 \pm 0.15$ & $8.2467 \pm 0.0093$ & $8.2446^{+0.0093}_{-0.0095}$ \\
$T$ (yr) & $16.0 \pm 0.02$ & $16.0455 \pm 0.0013$ & $16.0447 \pm {0.0013}$ \\
$t_P - 2018$ & $0.37897 \pm 0.01$ & $0.379 \pm 0.00016$ & $0.3780\pm{0.0002}$  \\
$a$ (mas) & $125.5 \pm 0.9$ & $125.058 \pm 0.041$ & $125.0936^{+0.0445}_{-0.0444}$ \\
$e$ & $0.8839 \pm 0.0019$ & $0.884649 \pm 0.000066$ & $0.8846\pm{0.0001}$ \\
$i \,(^\circ)$ & $134.18 \pm 0.4$ & $134.567 \pm 0.033$ & $134.6075^{+0.0285}_{-0.0297}$ \\
$\Omega \,(^\circ)$ & $226.94 \pm 0.6$ & $228.171 \pm 0.031$ & $228.1602 ^{+0.0325}_{-0.0311}$  \\
$\omega \,(^\circ)$ & $65.51 \pm 0.57$ & $66.263 \pm 0.031$ & $66.2578^{+0.0312}_{-0.0295}$ \\
$x_0$ (mas) & $-0.31 \pm 0.34$ & $-0.9 \pm 0.14$ & $0.0748^{+0.1313}_{-0.1342}$ \\
$y_0$ (mas) & $-1.29 \pm 0.44$ & $0.07 \pm 0.12$ & $0.1002^{+0.0941}_{-0.0897}$ \\
$v_{x,0}$ (mas/yr) & $0.078 \pm 0.037$ & $0.08 \pm 0.01$ & $0.0753^{+0.0138}_{-0.0139}$ \\
$v_{y,0}$ (mas/yr) & $0.126 \pm 0.047$ & $0.0341 \pm 0.0096$ & $0.0619^{+0.0177}_{-0.0185}$ \\
$v_{z,0}$ (km/s) & $8.9 \pm 4$ & $-1.6 \pm 1.4$ & $-2.2932^{+1.7327}_{-1.3011}$ \\
\hline
$\alpha $  & -- & -- & $2.2078^{+1.6717}_{-1.3203}$ \\
$\beta $  & -- & -- & $5.3230^{+2.8320}_{-1.6542}$ \\
$\gamma $  & -- & -- & $1.8376^{+0.0164}_{-0.0162}$ \\
$\rho_s \, (M_\odot/pc^3)$ & -- & -- & $0.0072^{+0.0008}_{-0.0009}$ \\
$r_s $ (kpc) & -- & -- & $14.8442^{+1.3116}_{-1.3541}$ \\
\hline
\end{tabular}
\caption{Median values and $1\sigma$ credible intervals of the MCMC samples for the parameters of the S2 orbital model. For comparison, we also show the best-fit parameter values from \cite{pdata1} and \cite{prior1}, which were obtained using Keplerian and post-Newtonian approaches, respectively, assuming the absence of a dark matter halo, whereas in this work we solve the full geodesic equations. The posterior distributions in the two reference works exhibit symmetric $1\sigma$ intervals, while our results show mild asymmetry.}
\label{fit}
\end{table*}

\begin{figure}[htbp]
    \centering
\includegraphics[scale=0.55]{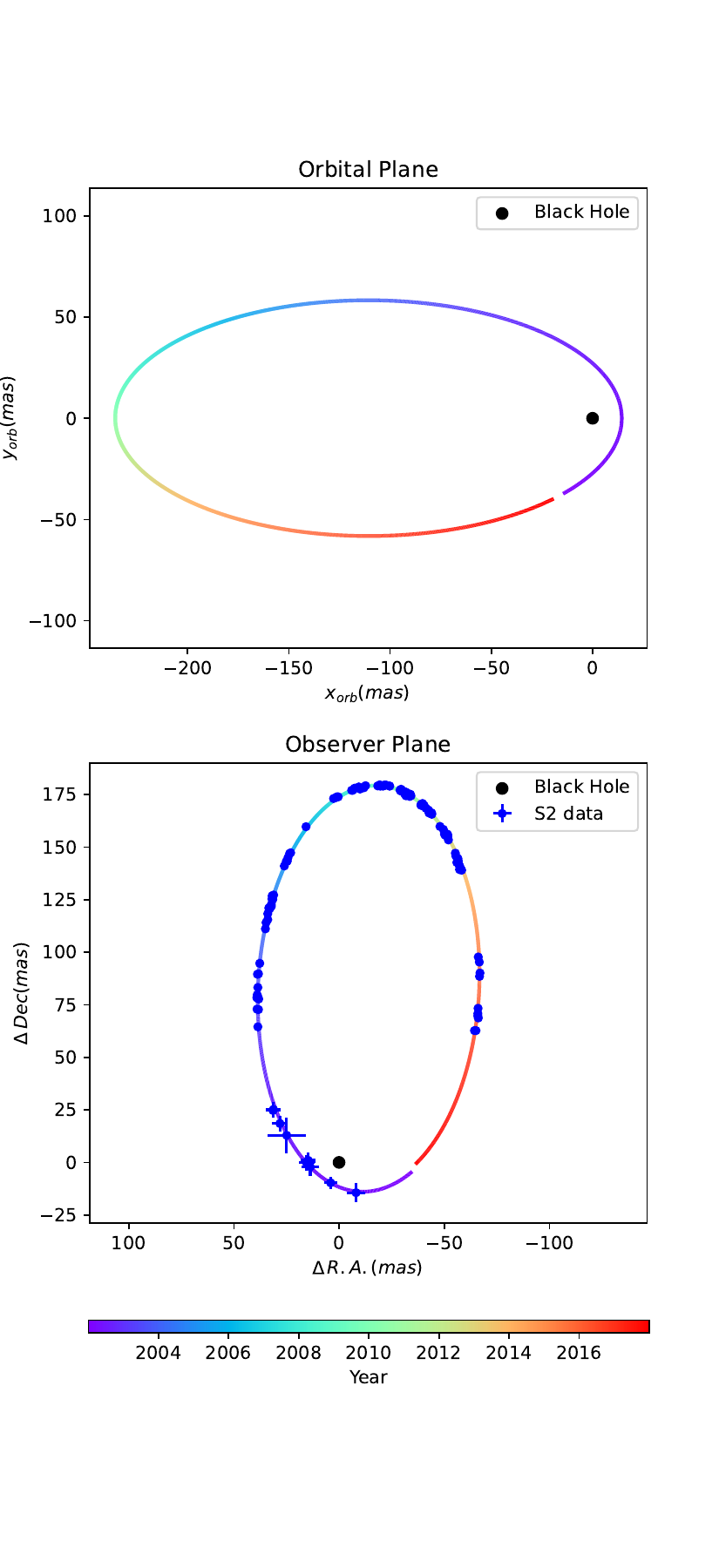}
\caption{Trajectories of the S2 star (colored curves) based on the best-fitting parameters listed in Table~\ref{fit}. The observational data of S2 are also shown. For clarity, only the orbit and data from 2002 to 2018 are displayed. The top panel presents the trajectories in the orbital plane, while the bottom panel shows the trajectories projected onto the observer plane. The rainbow color bar below indicates the corresponding years of the trajectories.}
\label{orbit}
\end{figure}

For clarity, we illustrate the S2 star trajectories in both the orbital plane and the observer plane in Fig.~\ref{orbit}, based on the best-fitting parameters listed in Table~\ref{fit}. The observational data of S2 are also shown. To enhance readability, only data and model trajectories from 2002 to 2018 are displayed. It is evident that the model provides an excellent fit to the observations, with the predicted trajectories passing through all data points within the corresponding error bars, thereby validating the dark matter halo-modified black hole model.

\begin{figure}[htbp]
    \centering
\includegraphics[scale=0.5]{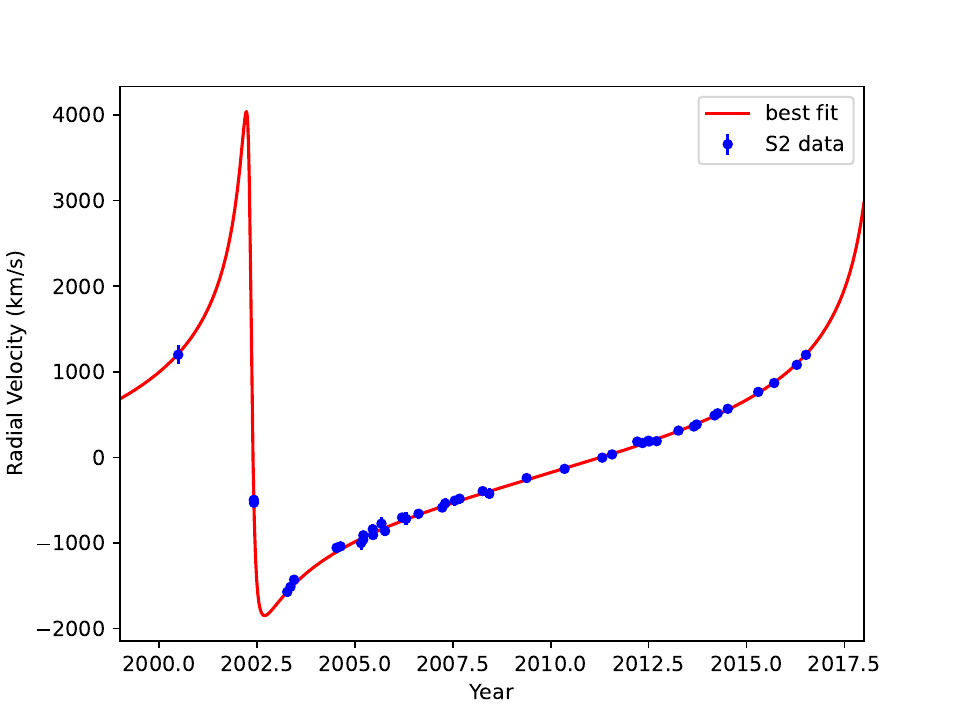}
\caption{Radial velocity of the S2 star (red curve) based on the best-fitting parameters listed in Table~\ref{fit}, along with the corresponding observational data. For clarity, only the data and model predictions from 1999 to 2018 are shown.}
    \label{vlo}
\end{figure}

In Fig.~\ref{vlo}, we present the radial velocity of the S2 star based on the best-fitting parameters listed in Table~\ref{fit}, along with the corresponding observational data. For clarity, only the data and model predictions from 1999 to 2018 are displayed. The radial velocity model curve provides an excellent fit to the observations, further confirming the validity of dark matter halo-modified black hole model.

\section{Conclusion}\label{sec5}

In this work, we have investigated the observational characteristics of the Galactic Center black hole, Sgr A$^*$, embedded in a general double power law dark matter halo, focusing in particular on the black hole shadow and the orbit of the S2 star. We constructed a modified black hole metric including the dark matter halo and studied its geodesics. We then explored how the halo parameters influence both the shadow and the S2 star orbit, and used observational data to constrain the parameter space of the dark matter halo. These results provide valuable insights into the properties of the central black hole and the surrounding dark matter distribution in the Milky Way.

For the black hole shadow, we computed the total number of photon orbits, light trajectories, and the resulting shadow radius. We found that the presence of dark matter generally enlarges the shadow radius. In particular, larger values of $\alpha$ and $\gamma$, and smaller values of $\beta$, lead to a greater shadow size. Furthermore, we mapped the two-dimensional parameter space for several parameter combinations and compared our results with the observational data of Sgr A$^*$ from the EHT. We found that larger $\beta$ together with smaller $\alpha$ and $\gamma$ correspond to broader parameter regions consistent with the observed shadow. Additionally, the parameters $\rho_s$ and $r_s$ are found to be negatively correlated within the allowed parameter space.

For the S2 star orbit, we jointly constrained the orbital parameters, zero-point offset and drift parameters, and dark matter halo parameters using long-term observations of the star's position and radial velocity through Bayesian inference and MCMC sampling. The posterior distributions and best-fit values for each parameter are presented and compared with previous results based on Keplerian and Post-Newtonian models. We also show the S2 star trajectories corresponding to the best-fit parameters in both the orbital plane and the observer plane (see Fig.~\ref{vlo}), demonstrating excellent agreement with the observational data.

As observational capabilities improve, more accurate and reliable data on the black hole shadow and S2 star orbit will become available. This will allow stronger and more precise constraints on the dark matter halo distribution in the Milky Way. The dark matter halo-modified black hole metric presented in (Eq.~\ref{fre}) and the corresponding observational constraints investigated in this work provide a theoretical and phenomenological foundation for future studies.

\section*{Acknowledgments}
Zhen Li acknowledges financial support from the Start-up Funds for Doctoral Talents at Jiangsu University of Science and Technology. 
This work is also supported by the National Natural Science Foundation of China (Grant No.~12505073). The author also thanks Xiao Yan Chew for helpful discussions.
\\

\end{document}